# Substituent-level Tuning of Frontier Orbital Energy Levels in Phthalocyanine/C$_{60}$ Donor-Acceptor Charge Transfer Pairs

Marco Olguin,[1] Luis Basurto,[1] Rajendra R. Zope,[2] and Tunna Baruah[1,2,a]

[1]Computational Science Program, The University of Texas at El Paso, El Paso, TX 79968, USA

[2]Department of Physics, The University of Texas at El Paso, El Paso, TX 79968, USA

[a]Author to whom correspondence should be addressed. Electronic mail: tbaruah@utep.edu

A fundamental requirement for donor-acceptor frontier orbital energy level offsets to achieve efficient charge photogeneration in organic photovoltaic systems is a relative energy ordering which creates a downhill energetic driving force conducive to the transferring of an electron from the donor molecule to the acceptor moeity. Since finding the ideal HOMO/LUMO level offset is not straightforward, it is of great importance to optimize the energy level positions of the donor and acceptor to allow efficient charge separation without losing the photo-generated voltage. An attractive quality in employing pi-conjugated macrocycles as chromophores in organic photovoltaic devices is the flexible incorporation of electron-rich and electron-deficient functional units along the conjugated periphery, where a variation of their strength allows for a systematic tuning of the molecular frontier orbital energy levels. The HOMO/LUMO level offset can be tailored by lowering the LUMO and/or raising the HOMO energy level of the molecule to obtain a minimum offset required to dissociate the Coulombically bound exciton at the donor-acceptor interface. This approach was applied in the sulfonation of the zinc-phthalocyanine (ZnPc) molecule at the ortho- and meta- positions of the macrocycle periphery with the objective of tuning the solubility and photocurrent properties of ZnPc/C60-based donor-acceptor organic solar cells, where altering the number of sulfonate substituent groups led to varying contributions to the device photocurrent. In the present study, we examine the effect of varying the number and position of sulfonate substituent groups attached to the ZnPc molecule on the frontier orbital energies of the donor and the CT excitation energy of the corresponding donor-acceptor complex. We have calculated several low-lying Charge Transfer (CT) excited-state energies for four non-covalently bound dyads composed of a sulfonated-ZnPc coupled to C60. Our results show that the di- and tri-sulfonated systems yield a CT state as the lowest-energy excited state in the system. In contrast, an energy re-ordering for the tetra-sulfonated ZnPc system leads to local excitations lying lower in energy than the CT state, displaying a possible deactivation pathway obstructing charge separation. Since several different donor-acceptor relative orientations may co-exist at an organic heterojunction, we compare the energetics of a few low-lying CT states for the end-on geometry of a di-sulfonated system to its co-facial orientation counterpart. The calculated CT excitation energies are larger for the end-on orientation in comparison to the co-facial structure by ~1.5 eV, which results principally from a substantial decrease in exciton binding energy in going from the co-facial to the end-on orientation. Furthermore, changes in relative donor-acceptor orientation have a larger impact on the CT energies than changes in donor-acceptor distance. TDDFT calculations on the various sulfonated ZnPc donor molecules show a significant splitting of the Q-band for only one of the four donor systems. Our present calculations, in line with previous experimental studies, show that the systematic variation of chemical functional groups is a promising avenue for the substituent-level tuning of various physical properties of organic semiconductors.

**INTRODUCTION**

The development of efficient photovoltaic devices derived from small-molecule organic semiconductors is driven, in part, by an extensive tunability of molecular properties afforded by the wide chemical functionality characteristic of substituted organic molecules. The broad and diversified group functionalization possible for organic molecules has lead to a search for new materials designed at the molecular level where improvements in efficiency may be realized by modifying chemical functional groups to alter the solubility, optical, electrical, and morphological properties of organic solar cells (OSCs).[1-15] The tuning of these properties in OSCs through chemical functionalization contributed to significant improvements in power conversion efficiency achieved within a decade with an increase from 1% to above 11%. Other advantages in developing OSCs as an efficient light-harvesting application for meeting increasing energy demands are a relatively simple synthesis, with great advances achieved in synthetic organic chemistry for pi-conjugated systems displaying attractive optoelectronic properties, and easy processability in manufacture.[1, 10, 16-33] Additionally, pi-conjugated systems are excellent sensitizers with good absorption coefficients in the visible part of the solar spectrum.[1, 2, 10, 34-43]

Two widely used deposition processes for organic molecular semiconductors are thermal deposition methods and solution processing techniques.[44, 45] Thermal vapor deposition processes allow for a highly reproducible thin film growth and for complete planar-heterojunction (PHJ) and bulk heterojunction (BHJ) solar cell device fabrication.[46-50] An important photomechanism in PHJ and BHJ organic solar cells originates at the donor-acceptor (D/A) interface, where charge transfer processes may yield a sought-after charge-separated state.[51] Whether an absorbed incoming photon is converted into a charge separated state and contributes to the photocurrent of the OSC by becoming mobile charge carriers becomes a complex multi-parameter problem which depends on several interrelated factors such as favorable HOMO/LUMO D/A energy differences needed to drive the charge separation process, the exciton diffusion length, and the morphology and phase separation of the active donor-acceptor layer.[51-58] The morphology impacts the overall solar cell performance in regard to charge transport properties, where the morphology of the active layer greatly influences the dissociation of Coulombic-bound excitons into charge carriers in the active bulk layer.[59-69] Within any D/A morphology, improved interfacial charge separation is achieved if molecular components are chosen such that an optimal energy offset between the LUMO of the donor and the LUMO of the acceptor is met.[57, 70-72]

As for solution-processed solar cell devices, a major challenge encountered is the synthesis of highly soluble donor and acceptor molecules that exhibit semiconductor properties, where many of the well-established molecular photo-sensitizers, such as phthalocyanine (Pc) and porphyrin molecules, exhibit a relatively low solubility and require environmentally inconvenient chlorinated solvents for solution processing.[73-87] The use of water as a solvent in solution-processed solar cells presents the advantage of simplifying the device fabrication process. Recently, Jones and co-workers employed water-soluble tetrasulfonated copper Pc (CuPc-S4) donor molecules in the production of solar cell devices where the active area was prepared from aqueous CuPc-S4 solutions and the corresponding acceptor layer consisted of $C_{60}$ fullerenes.[88] The devices showed a light-to-energy conversion of 0.32% under standard conditions, with no contribution to device photocurrent from the CuPc-S4 donor. Subsequently, Torres and co-workers set out to overcome the low photocurrent contribution of water-soluble sulfonated Pc donor systems by varying the number of sulfonate substituents at the periphery of the zinc-Pc macrocycle with the aim of shifting the donor-acceptor frontier orbital energies.[89]

Similar to the device performance obtained for the CuPc-S4/$C_{60}$ donor-acceptor pairs, the zinc analogue of the water-soluble sulfonated Pc donor component in the ZnPc-S4/$C_{60}$ system yielded a negligible contribution to the device photocurrent.[88, 89] Decreasing the number of sulfonate substituent groups for ZnPc resulted in an increasing contribution to the photocurrent, accompanied by a noticeable reduction in the open circuit voltage.[89] The lack of photocurrent exhibited by the device composed of ZnPc-S4/$C_{60}$ donor-acceptor pairs was attributed to a reduced free energy at the ZnPc-S4-$C_{60}$ interface, where the difference in donor and acceptor LUMO levels does not exceed the attractive Coulombic force of the photogenerated exciton necessary for charge separation at the interface. In turn, a comparison between the degree of sulfonation and the measured device $V_{OC}$ did not result in the expected pattern of values. If the device $V_{OC}$ is taken to be proportional to the difference in energy between the HOMO level of the donor and the LUMO level of the acceptor, the device exhibiting the lowest $V_{OC}$ among the set of ZnPc sulfonated systems should be the disubstituted ZnPc-S2 system. Experimental measurements show that the $V_{OC}$ is largest for ZnPc-S4 and lowest for ZnPc-S3 with ZnPc-S2 having an intermediate value.[89] Differences in morphology account for the irregular pattern of $V_{OC}$ values, where a marked difference in morphology between the ZnPc-S4 film and the di- and tri-sulfonated ZnPc films is revealed by atomic force microscopy.[89] Therefore, several loss processes related to smaller exciton diffusion lengths for the ZnPc-S2 and ZnPc-S3 systems may lead to the lower device Voc observed for ZnPc-S2/$C_{60}$ and ZnPc-S3/$C_{60}$.

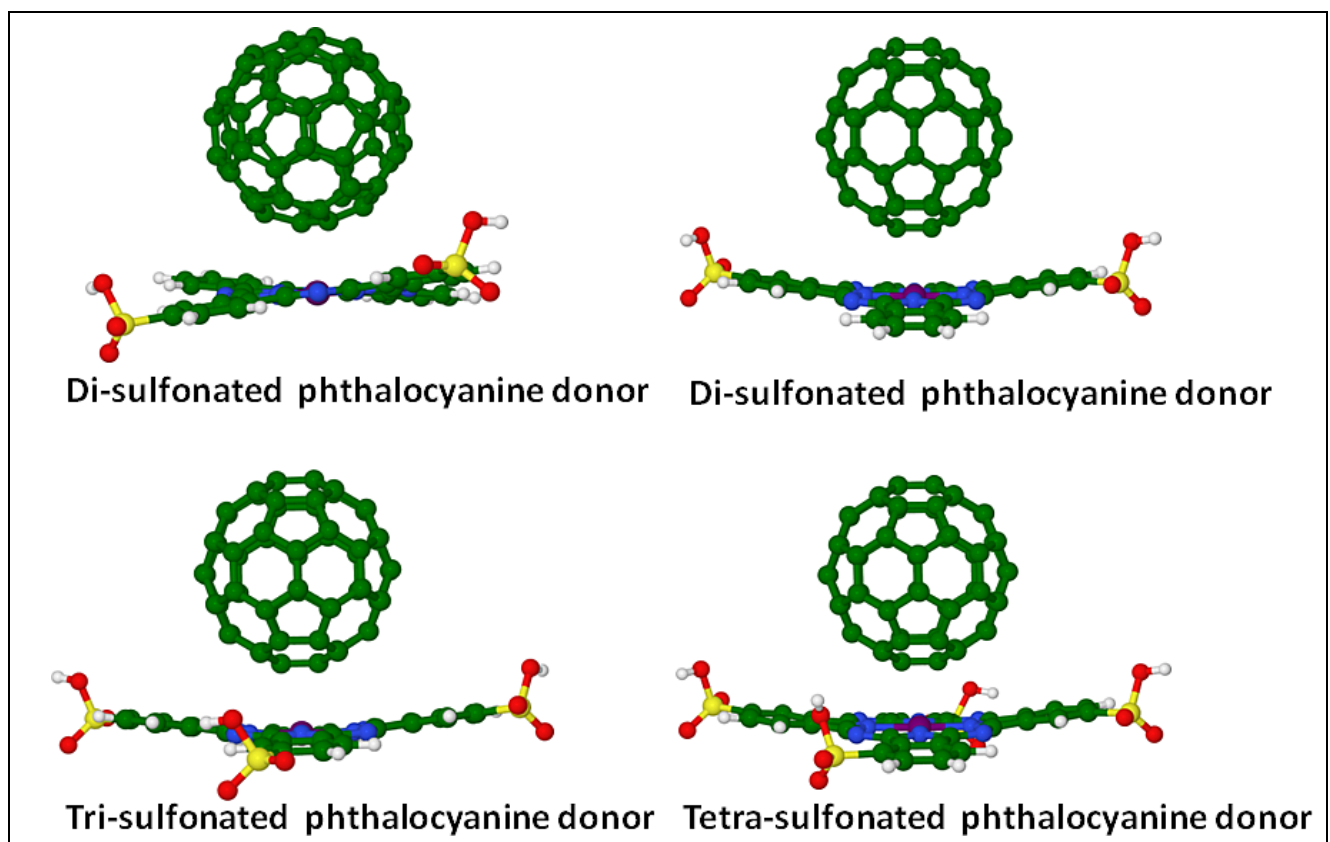

Figure 1. Four distinct sulfonated (zinc)phthalocyanine donor molecules coupled to $C_{60}$ acceptor.

In order to gain insight into the charge transfer processes for the ZnPcS/$C_{60}$ donor-acceptor pairs, we have performed ground-state and excited-state calculations on four different water-soluble sulfonated zinc-phthalocyanine (ZnPcS) donor molecules coupled to $C_{60}$ (figure 1). The four ZnPcS donor molecules (denoted as ZnPcS2A, ZnPcS2, ZnPcS3, and ZnPcS4) are shown in figure 2. From ground-state calculations of the ZnPcS donor and the $C_{60}$ fullerene acceptor in isolation and in complex, we can estimate the effect on the HOMO/LUMO energy level ordering of donor and acceptor component in forming a bound complex. In varying substituent groups on the donor molecules with the aim of shifting the frontier orbital energy level ordering and the CT excitation energies, the substituent type and position (ortho-, meta-) are important considerations. For instance, spectroscopic measurements indicate that electronegative substituent groups such as sulfonates shift the visible-spectrum absorption peaks (HOMO-to-LUMO and HOMO-to-LUMO+1 transitions) of ZnPc donor macrocycles. In addition, the different substituent positions shift the Q-band of the ZnPc macrocycle by varying degree. Therefore, varying the number of sulfonate substituents and substituent position in the ZnPc molecule will influence the frontier orbitals of the donor and the CT excitation energy of the donor-acceptor complex. In the present study, we have calculated several low-lying charge transfer excited-state energies for non-covalently bound ZnPcS/$C_{60}$ dyads in two different donor sulfonate substitution settings: (1) all sulfonate substituents at a meta-position and (2) sulfonate substituents at mixed meta- and ortho- positions.

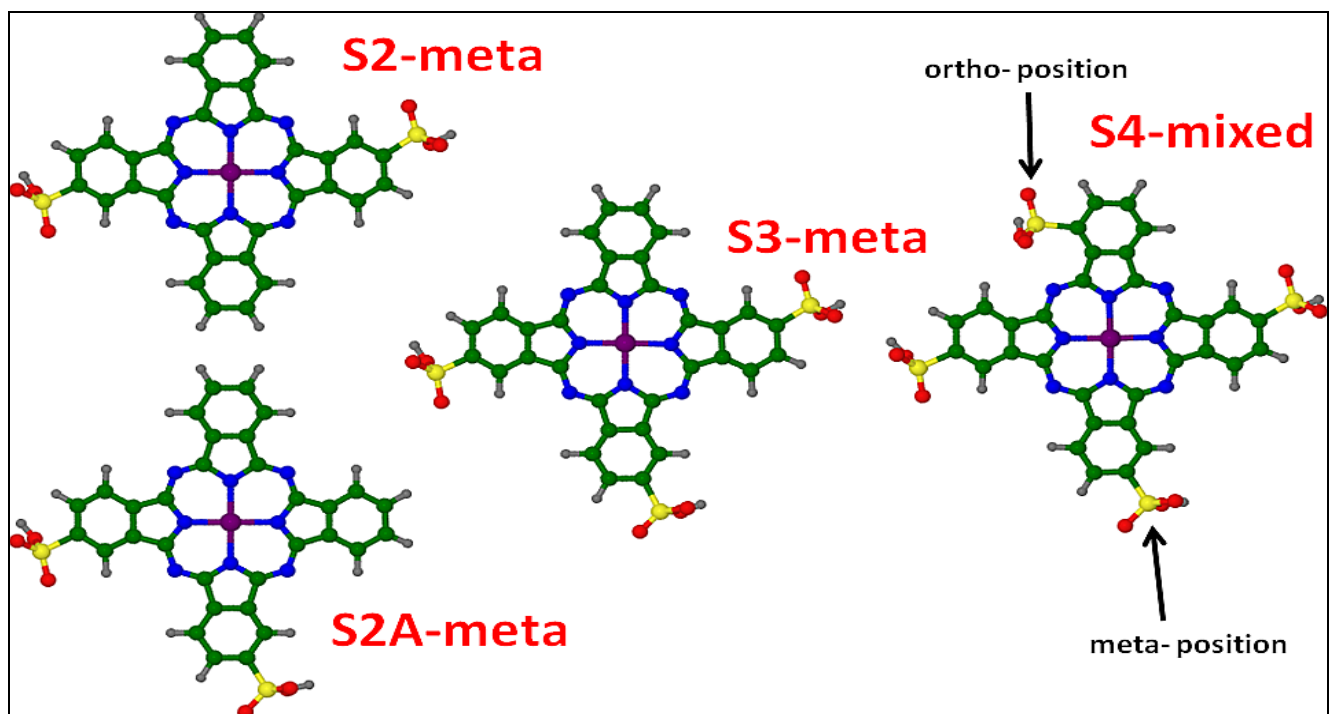

Figure 2. Four different water-soluble sulfonated (zinc)phthalocyanine molecules. The di-sulfonated moieties are denoted as S2 and S2A. The tri- and tetra- sulfonated compounds are labeled S3 and S4, respectively. Mixed ortho- and meta-substitution positions are shown for S4. The S2, S2A, and S3 molecules contain substituent groups strictly on meta-positions.

Several experimental and computational studies have addressed important concepts and developments in a current major area of investigation in organic photovoltaics that deals with understanding the physicochemical processes at the donor-acceptor interface. A complete description of the charge dissociation process in donor-acceptor based organic solar cells requires an understanding of the behavior and characteristics of various electronic processes and energy levels with respect to changes in donor-acceptor distance and relative orientation, electronic coupling, strength of the non-covalent interaction, and polarization effects arising from the donor-acceptor interface. In regard to the relative geometrical orientation between donor and acceptor component, two different orientations (shown in figure 3 as co-facial and end-on) for the dyad system were studied for the ZnPcS2A/$C_{60}$ donor-acceptor system. Since several donor-acceptor relative orientations may co-exist at a given planar- and bulk-heterojunction interface, studying both the co-facial and end-on configuration for the

ZnPcSA/$C_{60}$ dyad gives insight into the effect of geometrical orientation on the CT excitation energy. In going from the co-facial orientation to the end-on orientation, dispersion effects resulting from pi-pi stacking between the phthalocyanine macrocycle and the curved $C_{60}$ surface and other polarization effects will decrease. This change in polarization will influence the frontier orbital energy levels for each of the four donor-acceptor pairs. The difference in CT excitation energies between co-facial and end-on orientations will reflect the change in strength of the polarization effects and provide a reasonable estimate for the range in charge transfer excitation energies arising from several different co-existing relative dyad orientations at a donor-acceptor interface. In addition to varying donor-acceptor orientations present at the heterojunction interface, the intermolecular donor-acceptor distance will have a significant impact on the charge transfer energetics. Therefore, we have calculated the CT excitation energy as a function of donor-acceptor intermolecular distance spanning a range of 2.5 Å (figure 4). The intermolecular distance ($R_{co-facial}$ and $R_{end-on}$) CT calculations were performed for both the end-on and co-facial orientations of the ZnPcS2A/$C_{60}$ donor-acceptor pair. These calculations will provide a reasonable estimate of the range in CT excitation energies arising from the various donor-acceptor distances and orientations that will likely be present at the heterojunction interface of ZnPcS2A/$C_{60}$-based organic solar cells.

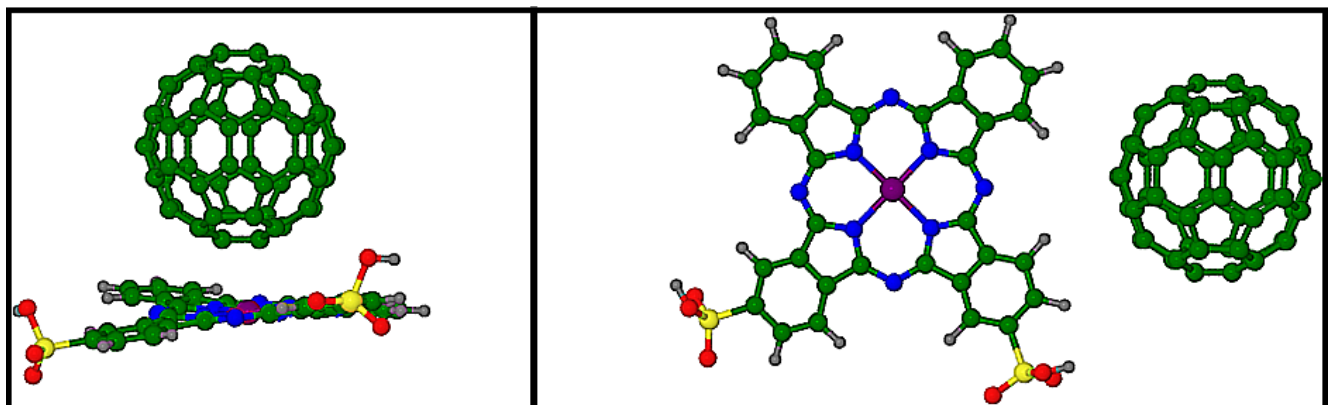

Figure 3. The left panel displays the co-facial orientation for the ZnPcS2A/$C_{60}$ dyad. The right panel shows the end-on orientation for the ZnPcS2A/$C_{60}$ dyad.

We also performed time-dependent Density Functional Theory (TDDFT) excited state calculations on each of the four different sulfonated zinc-phthalocyanine donor molecules in the various substituent patterns described above (figure 2). A strong absorption band (Q-band) resolved in a large number of phthalocyanine systems lies in the visible region at wavelengths between 650 nm to 670 nm, where symmetry plays an important role in determining the shape of the absorption peak for phthalocyanine-based macrocycles. Additionally, the substituent group type and the particular substituent ring positions influence the Q-band absorption energies. For instance, electron-withdrawing groups, such as sulfonyl and carboxyl groups, shift the Q-band to the red region of the visible spectrum. Group functionalization at the ortho- positions of the Pc macrocycle has a larger impact on the absorption spectra in comparison to similar meta- substituted complexes. The combination of ortho- and meta- substitution groups on the Pc macrocycle produces the largest Q-band shifts in the absorption spectra. Our results show that the di-sulfonated zinc-Pc molecules exhibit the largest red-shift relative to the non-substituted zinc-Pc molecule. We examine the effect of strict ortho- and meta- substitution and mixed ortho-/meta- substitution on the Q-band absorption (corresponding to HOMO-to-LUMO and HOMO-to-LUMO+1 transitions) of the sulfonated-ZnPc macrocycle chromophores.

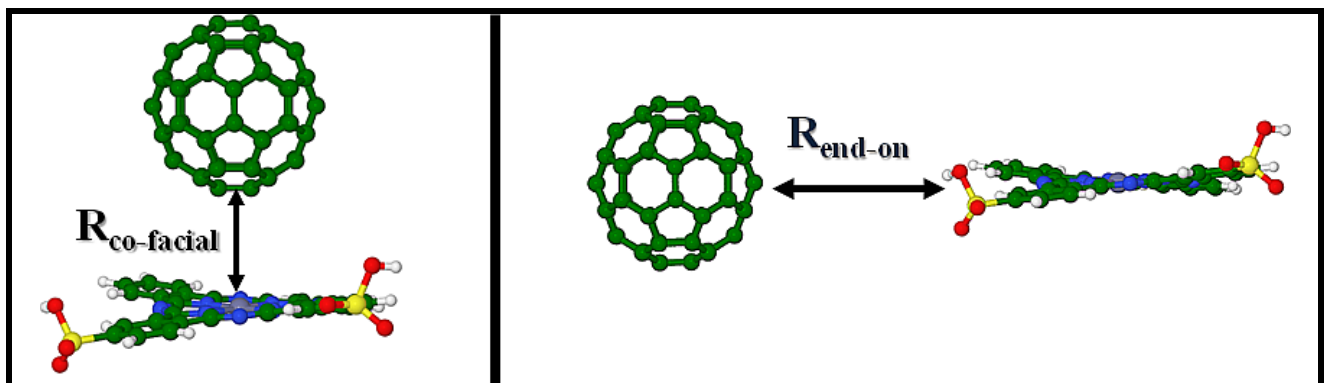

Figure 4. The left and right panels display the intermolecular distance scans ($R_{co\text{-}facial}$ and $R_{end\text{-}on}$) used in evaluating charge-transfer energies as a function of particle-hole distance.

## COMPUTATIONAL METHOD

The calculations reported here were carried out using Density Functional Theory (DFT) as implemented in the NRLMOL (Naval Research Laboratory Molecular Orbital Library) suite of codes. We employed the Perdew, Burke, and Ernzerhof (PBE) exchange-correlation energy functional within the generalized gradient approximation for all calculations reported here. The calculations were performed at the all-electron level using a large Gaussian basis set specially optimized for the PBE functional used in this work. The basis set for a given atom is contracted from the same set of primitive gaussians. The numbers of the primitive gaussians, s-type, p-type, and d-type contracted functions, along with the range of the exponents are given in Table I. This basis set resulted in a total of ~4420 basis functions for the ZnPcS/$C_{60}$ dyads studied here. All ground-state and excited-state calculations were performed using spin-polarized wavefunctions. The P-ΔSCF excited state DFT method has been implemented in the NRLMOL code and used here to determine the energies of the charge transfer excited state transitions. To obtain the excitation energy, an electron from an occupied state is placed in an unoccupied orbital and the self-consistent problem is solved using the perturbative ΔSCF method. The energy of the triplet state is obtained if the two unpaired electrons in the particle-hole state are of the same spin. However, if two unpaired electrons in the particle-hole state have opposite spin, then such a state is a mixed state (a 50–50 mixture of pure singlet and triplet states) with an energy that is an average of the singlet and triplet set. The energy of the singlet state may be calculated using the Ziegler-Rauk post-SCF spin-purification correction by subtracting the triplet energy from two times the energy of the mixed state. The P-ΔSCF method provides accurate estimates of the experimentally obtained charge transfer excited state energies for a set of 12 supramolecular Tetracyanoethylene (TCNE)-hydrocarbon dyads. Previously calculated CT excitation energies for porphyrin-$C_{60}$ co-facial dyads are in excellent agreement with the range of experimental values reported in the literature for similar porphyrin-fullerene systems. The method has also been applied to the study of charge transfer energetics in relation to varying geometrical orientation of the tetraphenyl-porphyrin/$C_{60}$ (TPP/$C_{60}$) and (zinc)tetraphenyl-porphyrin/$C_{60}$ (ZnTPP/$C_{60}$) supramolecular dyads. The excitation energy and oscillator strength calculations were carried out using time-dependent density functional response theory as implemented in the Gaussian09 program. The TDDFT calculations, carried out at the PBEPBE/6-311+G(d,p) optimized structures, were done using the same basis set as those used in the ground-state DFT calculations for the four ZnPcS donor molecules shown in figure 2. A previous TDDFT study on zinc-phthalocyanine showed that increasing the basis set size by adding diffuse functions and using larger triple-zeta basis sets had a small effect on the calculated excitation energies and oscillator strengths.

| Atom | s-type | p-type | d-type | Primitives | Exponent Range |
|------|--------|--------|--------|------------|----------------|
| C    | 5      | 4      | 3      | 12         | $2.22 \times 10^4 - 0.077$ |
| H    | 4      | 3      | 1      | 6          | $7.78 \times 10 - 0.075$ |
| N    | 5      | 4      | 3      | 13         | $5.18 \times 10^4 - 0.25$ |
| O    | 5      | 4      | 3      | 13         | $6.12 \times 10^4 - 0.10$ |
| S    | 6      | 5      | 3      | 17         | $6.72 \times 10^5 - 0.07$ |
| Zn   | 7      | 5      | 4      | 20         | $5.0 \times 10^6 - 0.055$ |

TABLE I. The numbers of s-, p-, and d-type contracted functions, number of primitive gaussians and the range of the gaussian exponents used for each atom.

## RESULTS AND DISCUSSION

## TDDFT CALCULATIONS

An important charactersitic of phthalocyanine macrocyles is its metallation capacity that spans ample chemical space from main group metals to transition metals and lanthanides to actinides. A range in the Q-band peak between 620 nm and 720 nm is observed for metallated Pc molecules with varying metal size, coordination, and oxidation state. Phthalocyanine molecules containing closed-shell metal atoms, such as zinc(II), show a maximum absorption peak near 670 nm. On the other hand, open-shell metal ions may interact strongly with the phthalocyanine ring and result in blue-shifted Q-bands with absorption maxima at around 630 nm to 650 nm. Metallated phthalocyanines adopt a higher symmetry than the corresponding free base phthalocyanine system with the incorporation of a metal ion inside the central cavity. This leads to a single peak Q-band in the visible range. The spectra of the corresponding free-base Pc molecules of lower symmetry yields a split Q-band. Also, substituent groups that preserve an overall symmetric metallated-phthalocyanine macrocyle exhibit only one absorption peak for the corresponding Q-band, whereas non-symmetric substitution breaks the molecular symmetry and gives rise to a split Q-band.

The spectral assignments of ZnPc (Q, B, N, L, and M bands) were first reported by Edward et al. based on broad gas-phase high-temperature spectra. Previous studies of the broad-range absorption spectrum for ZnPc using the time-dependent extension of DFT (TDDFT) gave results in excellent agreement with various experimental spectra. Since the zinc-phthalocyanine (ZnPc) macrocycle structure can be derived from the zinc-porphyrin (ZnP) molecular structure through combined tetraaza substitutions and tetrabenzo annulations, a particular TDDFT study of the combined effects of tetraaza and tetrabenzo groups on the structure and spectrum of ZnP showed a significant difference. Mainly, the near degeneracy of the HOMO and HOMO-1 orbitals of ZnP, which provides the basis for Gouterman's four-orbital model description of the frontier orbital transitions, breaks down for the ZnPc macrocycle. In zinc-phthalocyanine (ZnPc), the near degeneracy of the $a_{2u}$ orbitals with other occupied orbitals gives rise to a complex structure in the higher energy regions of the spectra. Consequently, the orbitals lying lower in energy than the HOMO level are found to be well (2.57 eV) separated from the HOMO, where the HOMO and lower occupied orbitals of ZnPc all have a significant pi contribution from the benzo rings. For orbitals that are known to give rise to specific well-characterized transitions such as the Q-band absorption peak(s), the energy shifts are useful in interpreting the variation in excitation energies and intensity produced by different substituents.

Our TDDFT calculation at the PBEPBE/6-311+G(d,p) level of theory for zinc-phthalocyanine gave a Q-band absorption peak value of 1.88 eV in excellent agreement with the experimentally determined value of 1.89 eV (gas-phase spectrum). This level of theory was used to calculate the near

Q-band absorption spectrum of all sulfonated-phthalocyanine donor molecules shown in figure 2. The TDDFT calculations show that the absorption peak for one of the two disulfonated zinc-phthalocyanine molecule conformers (ZnPcS2 shown in the left-side of figure 5) yields a split Q-band of 0.12 eV with respect to the non-sulfonated zinc-phthalocyanine molecule. Interestingly, the HOMO-to-LUMO transition (661 nm) is red-shifted with respect to the calculated single-peak value of ZnPc (639 nm), whereas the HOMO-to-LUMO+1 transition (621 nm) exhibits a blue-shift. In contrast, the other di-sulfonated isomer (ZnPcS2A shown in the right-side of figure 5) displays a negligible change in absorption shape and magnitude in comparison to the single-peaked Q-band absorption of the ZnPc macrocycle. Table I shows the calculated Q-band absorption values for each of the four sulfonated donor molecules (ZnPcS2A, ZnPcS2, ZnPcS3, and ZnPcS4) and the non-sulfonated ZnPc molecule. For the tetra-sulfonated donor molecule (ZnPcS4), a molecule which contains sulfonate substituents at the meta- and ortho- positions, the Q-band exhibits a split similar to the ZnPcS2 donor molecule but smaller in magnitude. The tri-sulfonated system (ZnPcS3) shows a small (~0.03 eV) red-shift for both the absorption transition peaks originating from the HOMO orbital.

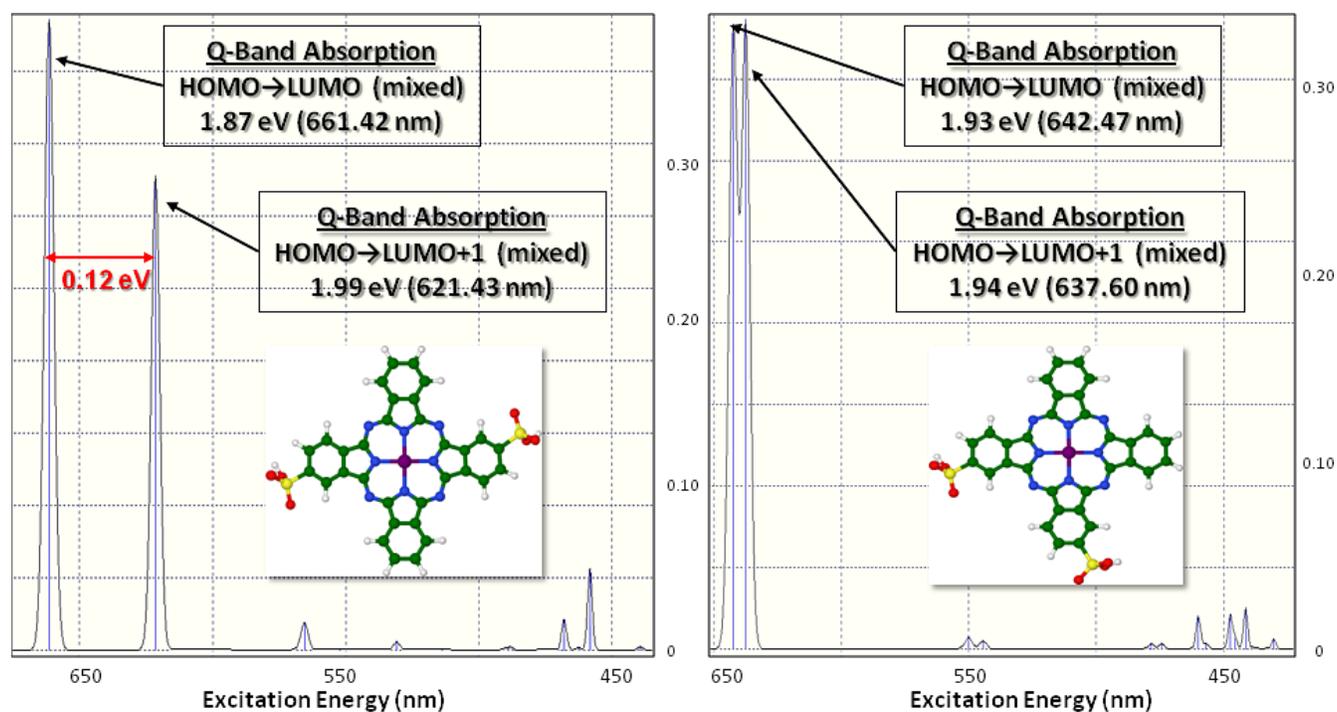

Figure 5. The split Q-band absorption peaks for the di-sulfonated ZnPcS2 donor molecule.

| Q-Band Absorption (nm) | ZnPc Base | ZnPcS2A | ZnPcS2 | ZnPcS3 | ZnPcS4 |
|---|---|---|---|---|---|
| HOMO → LUMO (mixed) | 639.67 | 642.47 | 661.42 | 657.26 | 643.75 |
| HOMO → LUMO+1 (mixed) | 638.63 | 637.60 | 621.43 | 646.46 | 633.18 |

Table I. Transitions calculated by the TDDFT method at the PBEPBE/6-311+G(d,p) level of theory.

# GROUND STATE CALCULATIONS

For the four distinct sulfonated ZnPc molecules, upon forming a complex with the $C_{60}$ fullerene, the HOMO orbital becomes the HOMO orbital of the phthalocyanine-fullerene complex. However, increasing the degree of sulfonation in the Zn-phthalocynanine macrocycle reduces the mismatch in energy between the HOMO level of the isolated ZnPcS-donor and the HOMO levels of the isolated $C_{60}$-acceptor, where the energy difference between the tetra-sulfonated ZnPc HOMO level and the $C_{60}$ 5-fold degenerate $h_u$ HOMOs in the complex is less than 0.1 eV. The five-fold degenerate HOMO orbitals for the isolated $C_{60}$ fullerene form the HOMO-1 to HOMO-5 orbitals in the complexes with each of the ZnPcS donor molecules. From the DFT ground-state calculations for each of the four co-facial ZnPcS/$C_{60}$ non-covalently bound dyads, figures 6 and 7 show the corresponding orbital energy diagrams with the orbital range spanning from HOMO-5 to LUMO+5 for both donor and acceptor molecule in isolation and in complex. Figure 6 displays a side-by-side comparison of the energy level ordering between a di-sulfonated ZnPc donor molecule (ZnPcS2) and the tri-sulfonated ZnPc (ZnPcS3) molecule and figure 7 compares the isomeric di-sulfonated ZnPc (ZnPcS2A) moiety with the tetra-sulfonated ZnPc (ZnPcS4) donor molecule. Since the energy ordering is similar for both di-sulfonated ZnPc systems, the comparisons emphasize the shift in the frontier orbital energy levels of the sulfonated ZnPc molecules in increasing the number of sulfonate substituent groups from two (ZnPcS2A and ZnPcS2) to three (ZnPcS3) and four (ZnPcS4) groups.

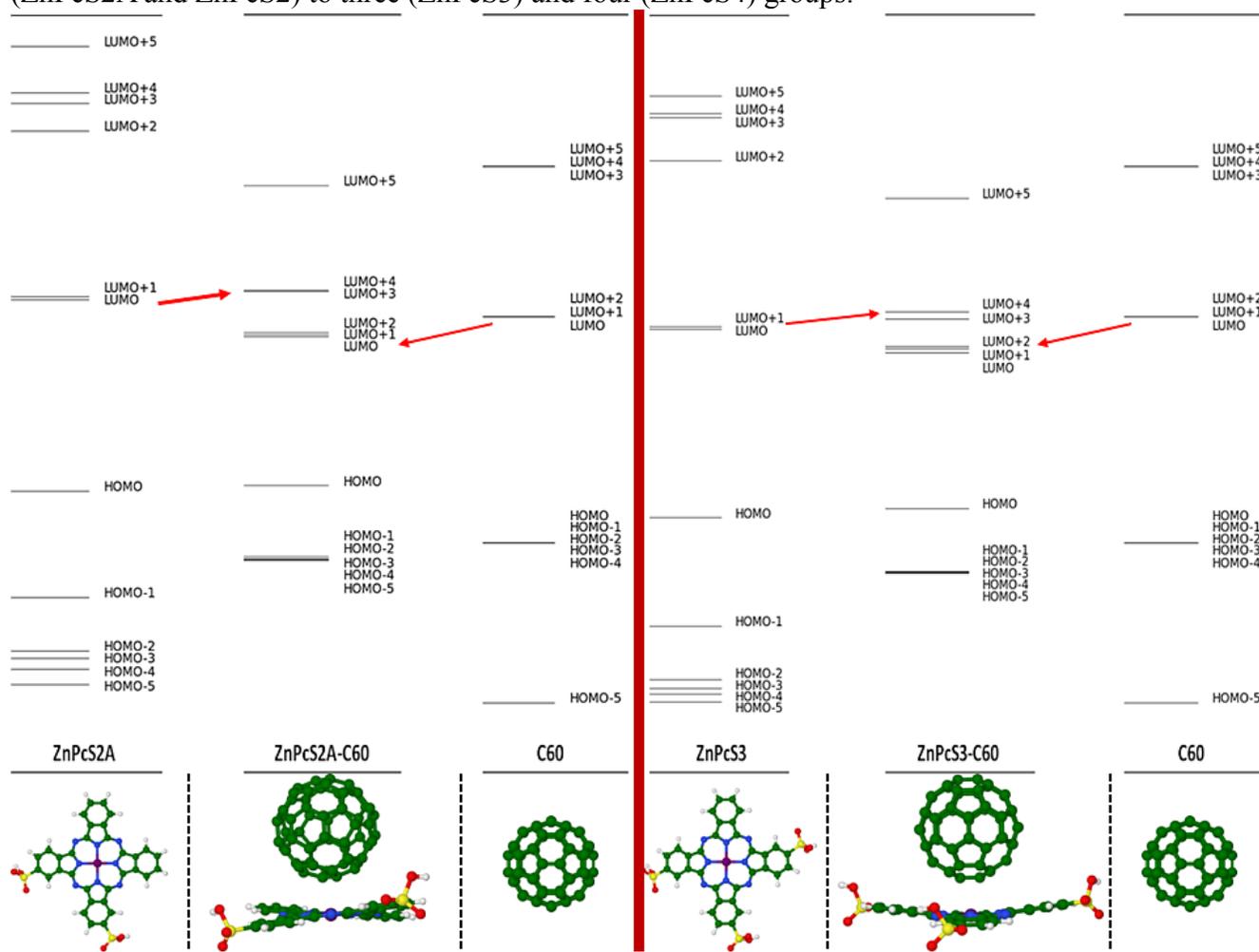

Figure 6. The left panel displays the orbital energy level ordering for the donor (left) and acceptor (right) molecule in isolation and in complex (middle) for the ZnPcS2A/$C_{60}$ system. Likewise, the right panel shows the energy ordering for the ZnPcS3/$C_{60}$ system.

The most important feature in the comparison of the energy level diagrams in regard to the frontier orbital energies known to impact the photovoltaic properties of a D/A-based organic solar cell is that both the HOMO and LUMO/LUMO+1 energy levels of the donor molecule in isolation incur a substantial lowering in energy in going from the di-sulfonated ZnPc system to the tri- and tetra-sulfonated ZnPc molecules. In both the isolated ZnPcS3 and ZnPcS4 ground-state systems, the LUMO and LUMO+1 orbitals are lower in energy than the 3-fold degenerate isolated-$C_{60}$ fullerene LUMO level. Moreover, the energy of the HOMO orbital for both the ZnPcS3 and ZnPcS4 molecules is very close to the energy of the 5-fold degenerate $C_{60}$ HOMO level. Despite a seemingly unfavorable donor/acceptor HOMO/LUMO energy level offset between the ZnPcS3/ZnPcS4 and $C_{60}$ molecular systems, in forming a bound phthalocyanine-fullerene complex the triply degenerate LUMO levels of the isolated $C_{60}$ molecule become the lowest lying LUMO orbitals of the complex in the ground state (as indicated by red arrows in figures 6 and 7). The LUMO/LUMO+1 orbitals of the sulfonated (zinc)phthalocyanine donor molecules become the higher-lying LUMO+3/LUMO+4 orbitals in the donor/acceptor complex.

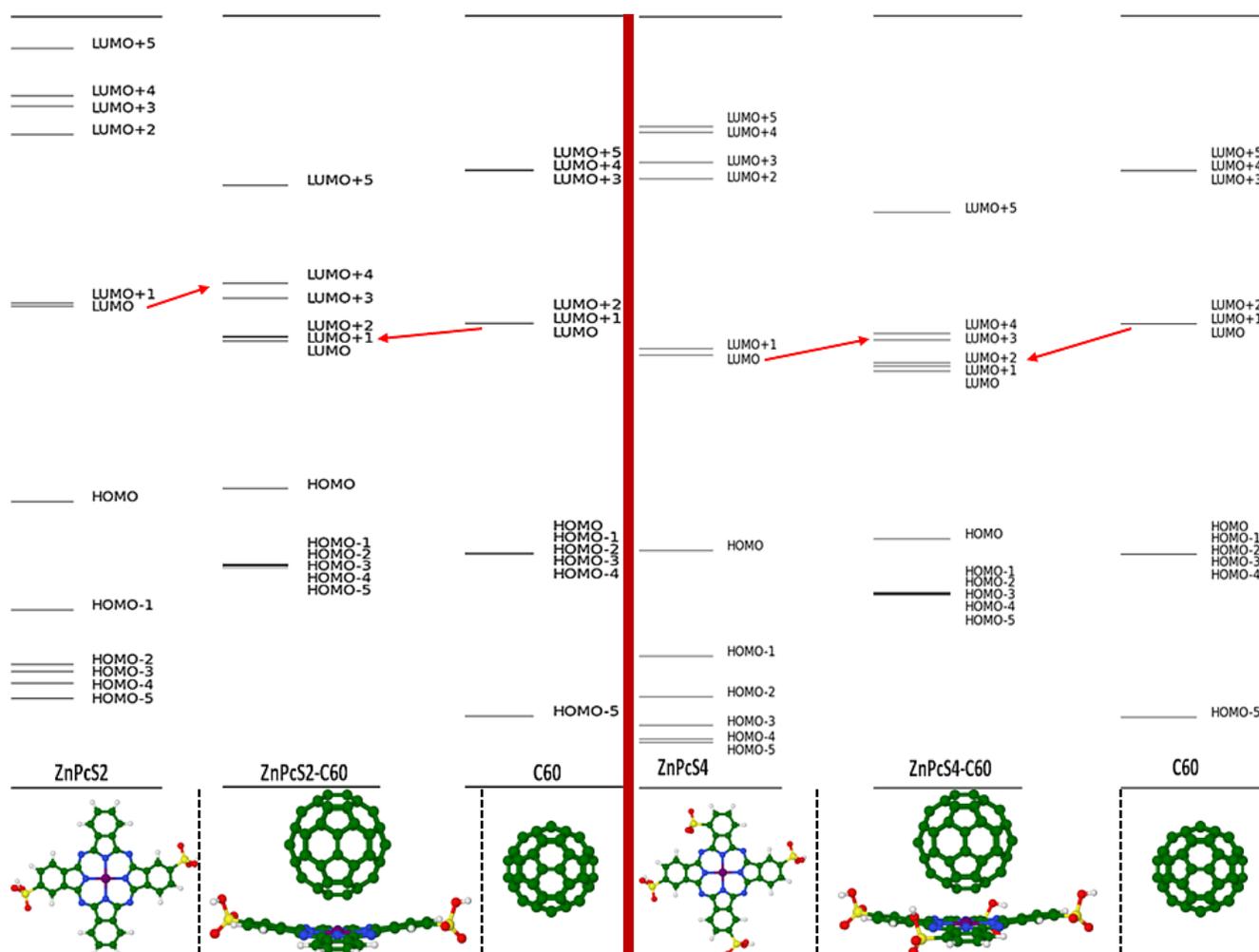

Figure 7. The left panel displays the orbital energy level ordering for the donor and acceptor molecule in isolation and in complex for the ZnPcS2/$C_{60}$ system. Likewise, the right panel shows the energy ordering for the ZnPcS4/$C_{60}$ system.

In Table II we show the DFT calculated ionization potential of the donor and electron affinity of the acceptor in isolation and in complexes. For a neutral molecule with N electrons, the ionization

potential (IP) can be calculated from the following expression:

$$IP = E(N-1) - E(N)$$

where E(N) is the self-consistent total energy of the molecule in the ground state and E(N-1) is the total energy of the cation. If the ionization process is rapid with respect to structural relaxation, then the energy of the cation E(N-1) may be calculated using the geometry of the neutral system and the resultant ionization energy is called the vertical IP. Similarly, the electron affinity (EA) can be calculated using the following definition:

$$EA = E(N) - E(N+1)$$

where E(N+1) is the self-consistent total energy of the anion calculated using the geometry of the neutral system. The IP and EA values reported in Table II were calculated within this ΔSCF scheme.

|       | S2   | S2A  | S3   | S4   | $C_{60}$ | $S2/C_{60}$ | $S2A/C_{60}$ | $S3/C_{60}$ | $S4/C_{60}$ |
|-------|------|------|------|------|----------|-------------|--------------|-------------|-------------|
| IP    | 6.79 | 6.82 | 7.00 | 7.14 |          | 6.59        | 6.64         | 6.77        | 6.90        |
| EA    |      |      |      |      | 2.67     | 3.13        | 3.13         | 3.28        | 3.37        |
| IP-EA |      |      |      |      |          | 3.46        | 3.51         | 3.49        | 3.52        |
| EBE   |      |      |      |      |          | 2.05        | 2.00         | 2.04        | 1.95        |

Table II. The ionization potential (IP) for each of the four donor molecules and the electron affinity (EA) of the $C_{60}$ acceptor molecule in isolation and in complex. The IP-EA value gives the quasi-particle gap. The exciton binding energy (EBE) is given.

The table shows that the IP value increases with the number of added sulfonate substituent groups to the zinc-phthalocyanine macrocycle. This is reflected in the energy level diagrams given in figures 6 and 7, where the HOMO level of the donor component is lower in energy for the tri- and tetra-sulfonated ZnPc molecules (ZnPcS3 and ZnPcS4; right-panel of figures 6 and 7) in comparison to the di-sulfonated systems (ZnPcS2 and ZnPcS2A; left panel of figures 6 and 7). In forming a complex with the $C_{60}$ fullerene, the four sulfonated-ZnPc systems (ZnPcS2, ZnPcS2A, ZnPcS3, and ZnPcS4) exhibit a lowering in energy of the IP level by approximately 0.2 eV. The EA level of the acceptor component is shifted upward in energy in forming a complex with the ZnPcS donor molecules, where a larger shift takes place for the ZnPcS3/$C_{60}$ and ZnPcS4/$C_{60}$ dyads with a significant change of ~0.7 eV. In figure 8 we show an isosurface of the difference between the ground state electron density of the dyad at equilibrium separation and the densities of the isolated components. The blue (grey) surface shows the region where the density difference is negative (positive). From the figure it is seen that substantial charge redistribution takes place on the curved fullerene surface near the interface with the semi-planar phthalocyanine macrocycle plane. The polarization of the fullerene thus contributes to the formation of the interfacial dipole. Overall, the quasi-particle gap, calculated as the $IP_{dyad}$-$EA_{dyad}$, and the exciton binding energy (EBE) are similar in value for the group of four ZnPcS/$C_{60}$ dyads (Table II). The exciton binding energy (EBE) was calculated using the quasi-particle gap and the lowest charge transfer excitation energy ($CT_{lowest}$) according to the following equation:

$$EBE = (\text{quasi-particle gap}) - CT_{lowest} = (IP-EA)_{dyad} - CT_{lowest}$$

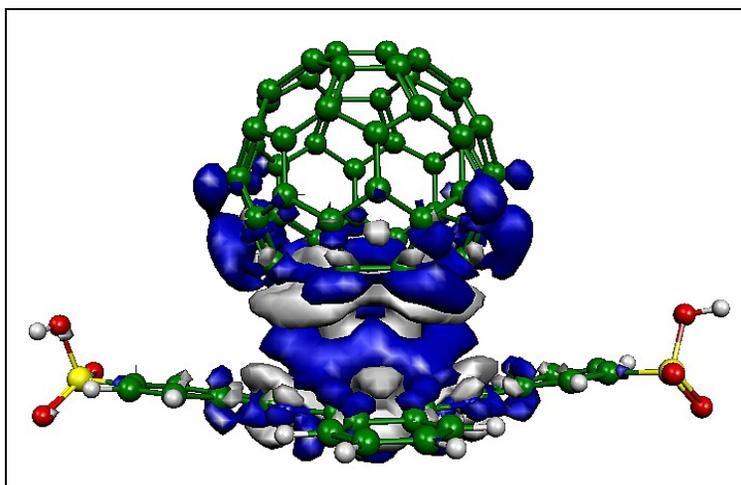

Figure 8. The difference in ground state density of the ZnPcS2/$C_{60}$ dyad and its isolated components.

## CHARGE TRANSFER EXCITED STATE CALCULATIONS

By calculating several low-lying charge transfer excited-state energies for the four different non-covalently bound dyads, we examine the relationship between the degree of sulfonation for the donor component of the charge transfer complex and the donor-acceptor charge transfer excitation energy. This understanding is important from the perspective of device performance efficiency, where the device $V_{OC}$ is largely determined by the CT excitation energy. The power conversion efficiency of a donor-acceptor based organic solar cell is directly proportional to the $V_{OC}$ as shown in equation 1:

$$\text{Power Conversion Efficiency: } \eta = V_{OC} J_{SC} FF / P_{in} \quad (1)$$

Equation (1) shows three widely used parameters in the characterization of a solar cell: the short-circuit current density ($J_{SC}$), the open-circuit voltage ($V_{OC}$), and the fill factor (FF). $J_{SC}$ is a measure of the effectiveness of the organic semiconductor in converting absorbed photons into charge carriers and the FF term describes the quality of the solar cell in terms of photogenerated charge carriers reaching the electrodes. In regard to the $V_{OC}$, a dependence on the energy ordering of the donor-acceptor frontier orbitals allows for the tuning of $V_{OC}$ by combining various donor and acceptor molecules with HOMO-LUMO energy differences within a targeted range. Another way of tuning the energy levels of the frontier orbitals is by controlling the relative orientation between the donor and acceptor molecule. In using group functionalization of organic molecules to create increasingly favorable donor-acceptor energy band offsets, accurate CT excited state calculations complement the experimental molecular tuning of important photovoltaic properties by giving insight into the effect of varying donor substituent groups and varying donor-acceptor distance/orientation on the CT energetics of donor-acceptor complexes.

Figure 9 depicts the lowest few CT excited state transitions for the di-sulfonated phthalocyanine (ZnPcS2A) HOMO orbital to the $C_{60}$ LUMOs for the co-facial ZnPcS2A/$C_{60}$ complex by arrows originating from the donor states shown on the left side to the corresponding $C_{60}$ acceptor states displayed on the right side of the figure. The calculated singlet and triplet excitation energies for the CT excited state transitions between the ZnPcS-localized donor states and the $C_{60}$-localized acceptor states are displayed in figure 10. The singlet excitation energies are calculated following the prescription given by Ziegler et al., that is, by subtracting the triplet energy from twice the energy of the mixed state. A comparison between the singlet and triplet excitation energies is made side by side in figure 10 for each of the four different co-facial ZnPcS/$C_{60}$ dyads (ZnPcS2/$C_{60}$, ZnPcS2A/$C_{60}$, ZnPcS3/$C_{60}$, and ZnPcS4/$C_{60}$). The labeled Charge Transfer (CT) and Local Excitation (PE and FE) transitions are assignments to the di-sulfonated ZnPcS2A/$C_{60}$ and ZnPcS2/$C_{60}$ dyad systems, where the HOMO→LUMO, HOMO→LUMO+1, and HOMO→LUMO+2 excited state transitions correspond to donor-acceptor charge-transfer states. In turn, the HOMO→LUMO+3 and HOMO→LUMO+4 transitions correspond to local sulfonated-(zinc)phthalocyanine donor excitations for the ZnPcS2A/$C_{60}$ and ZnPcS2A/$C_{60}$ complexes. Our calculations give a lowest CT excited state lower in energy than the lowest-lying local donor excitations for the di-sulfonated systems, which may serve as an energy-level setting conducive to CT excited state transitions. In contrast, an energy level re-ordering (reversal) takes place among the frontier orbitals of the tetra-sulfonated dyad system (ZnPcS4/$C_{60}$), leading to the local donor excitation (1.53 eV) lying lower in energy than the ZnPcS4/$C_{60}$ CT band (1.57 eV). Thus, the CT excited state is not energetically favored in comparison to competing local donor excitation transition pathways for the ZnPcS4/$C_{60}$ donor-acceptor pair. Our excited state calculation results are in line with previous experimental studies on sulfonated (zinc)-phthalocyanine-$C_{60}$ molecular dyads employed in solar cells, and the related sulfonated (copper)-phthalocyanine-$C_{60}$ dyads, which show a negligible contribution to device photocurrent from the tetra-sulfonated systems and a corresponding improvement in device performance for di- and tri-sulfonated systems. Thus, the molecular tuning of CT states to obtain increasingly favorable frontier orbital energy offsets in donor/acceptor systems can be achieved through a systematic and judicious use of group functionalization of well-established pi-conjugated organic chromophore molecules such as porphyrin- and phthalocyanine-based macrocycles.

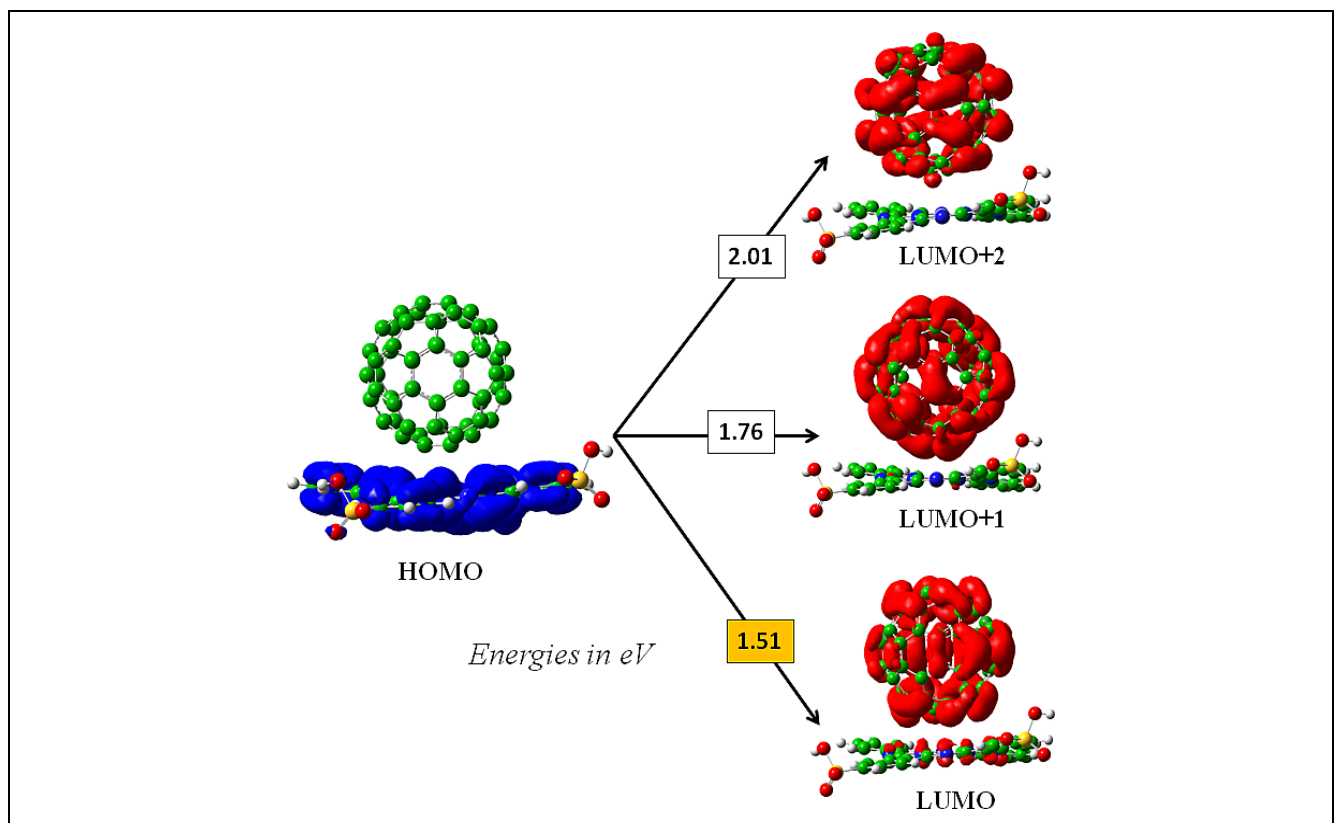

Figure 9. CT excited state transitions (in eV) for co-facial ZnPcS2A/$C_{60}$. Transitions are depicted by arrows originating from the ZnPcS2A donor state shown on the left (blue) to the corresponding $C_{60}$ acceptor states (red) displayed on the right.

The relative donor-acceptor orientation plays an important role in determining photophysical properties of charge transfer complexes. Several experimental and theoretical studies have shown that the relative position of the donor and acceptor components significantly impacts the interfacial electronic processes in organic solar cells. The different possible orientations at a heterojunction interface of phthalocyanine-fullerene supramolecular dyads will generate varied associated local electric fields which influence the charge transfer energetics. Additionally, the strength of the stabilizing non-covalent pi-pi interaction between a (zinc)phthalocyanine-$C_{60}$ dyad is expected to be maximal for the co-facial orientation, where the donor-acceptor surface-to-surface interaction is largest, and minimal for the end-on configuration. Dispersive-related polarization effects originating from the interaction between the sulfonated (zinc)phthalocyanine and fullerene charge distributions will also decrease in going from the co-facial orientation to the end-on configuration. Thus, calculating the charge transfer excitation energies for both the co-facial and end-on ZnPcS2A/$C_{60}$ supramolecular dyad may provide a reliable estimate for a range of achievable open-circuit voltage in a donor-acceptor based photovoltaic cell. In figure 11, we show the calculated excitation energy for the three lowest-lying CT states of the ZnPcS2A/$C_{60}$ molecular dyad in an end-on orientation. The calculated lowest-lying CT excitation energies are larger for the end-on orientation in comparison to the co-facial structure by ~1.5 eV, which primarily occurs due to a decrease in exciton binding energy in going from the co-facial to the end-on orientation, where the calculated exciton binding energy (EBE) for the ZnPcS2A/$C_{60}$ end-on orientation (0.46 eV) is much smaller than the EBE of its counterpart co-facial ZnPcS2A/$C_{60}$ orientation (2.00 eV). A comparison between the CT energetics of the di-sulfonated (zinc)phthalocyanine/$C_{60}$ (ZnPcS2A/$C_{60}$) dyad and the (zinc)tetraphenyl-porphyrin/$C_{60}$ (ZnTPP/$C_{60}$) dyad shows that the exciton binding energy difference in going from a co-facial orientation to an end-on orientation is two times larger for the ZnPcS2A/$C_{60}$ complex (1.54 eV) than the ZnTPP/$C_{60}$ system (0.70 eV).

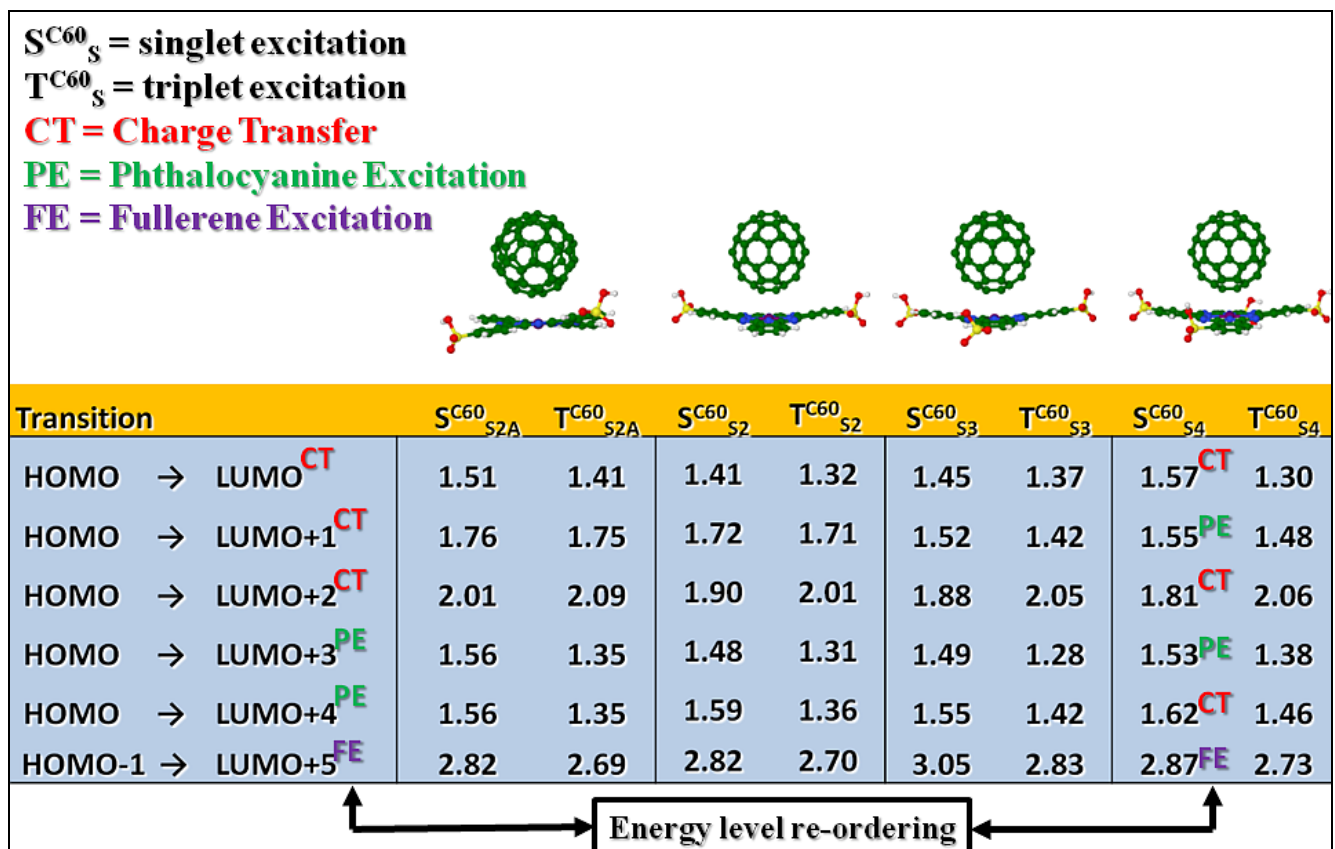

Figure 10. Calculated energies (in eV) for the charge transfer excited state transitions between the ZnPcS-localized donor states and the $C_{60}$-localized acceptor states. The HOMO state is localized on the ZnPcS component. The LUMO to LUMO+2 states are localized on the $C_{60}$ fullerene.

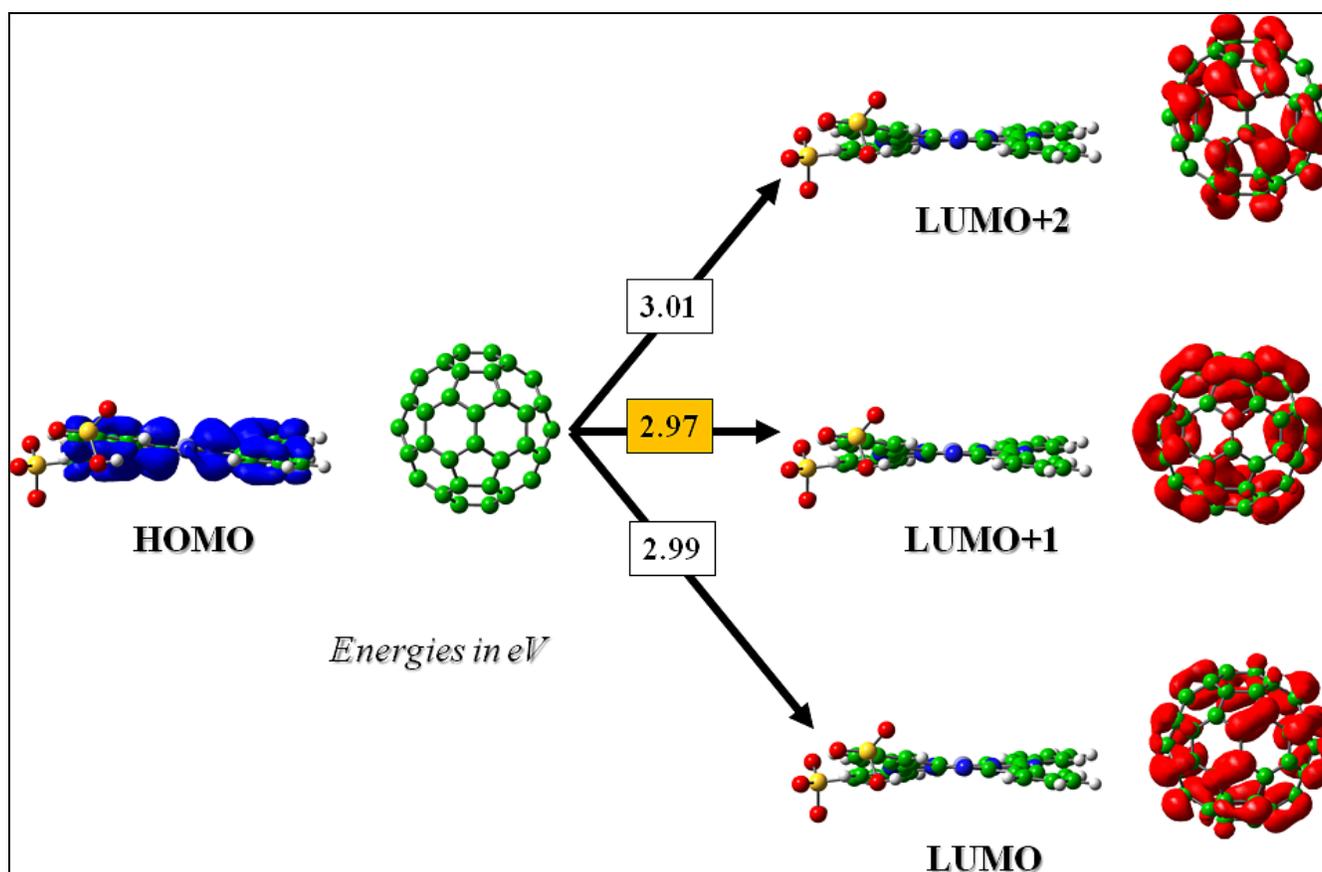

Figure 11. CT excited state transitions (in eV) for end-on ZnPcS2A/$C_{60}$. Transitions are depicted by arrows originating from the ZnPcS2A donor state shown on the left (blue) to the corresponding $C_{60}$ acceptor states (red) displayed on the right.

We further studied the effect on the CT excitation energy of varying the phthalocyanine-fullerene center-to-center distance for both the co-facial and end-on orientation of ZnPcS2A/$C_{60}$. In order to examine the behavior of the CT excitation energy as a function of particle-hole distance, we have calculated the lowest HOMO to LUMO CT energies of the ZnPcS2A/$C_{60}$ dyad for both co-facial and end-on orientations as a function of the ZnPcS2A-fullerene intermolecular distance spanning a center-to-center distance of 2.5 Å in five increments of 0.5 Å (figure 12). The calculated range in CT energy for the end-on intermolecular distance scan of the ZnPcS2A/C60 dyad is 0.27 eV, whereas the range in CT energy for the co-facial intermolecular distance scan is 1.16 eV. A comparison between the CT energy distance-scan of the di-sulfonated (zinc)phthalocyanine/$C_{60}$ (ZnPcS2A/$C_{60}$) dyad and the (zinc)tetraphenyl-porphyrin/$C_{60}$ (ZnTPP/$C_{60}$) dyad shows that the range of CT excitation energies is two times larger for the ZnPcS2A/$C_{60}$ complex (1.16 eV) than the ZnTPP/$C_{60}$ system (0.60 eV).

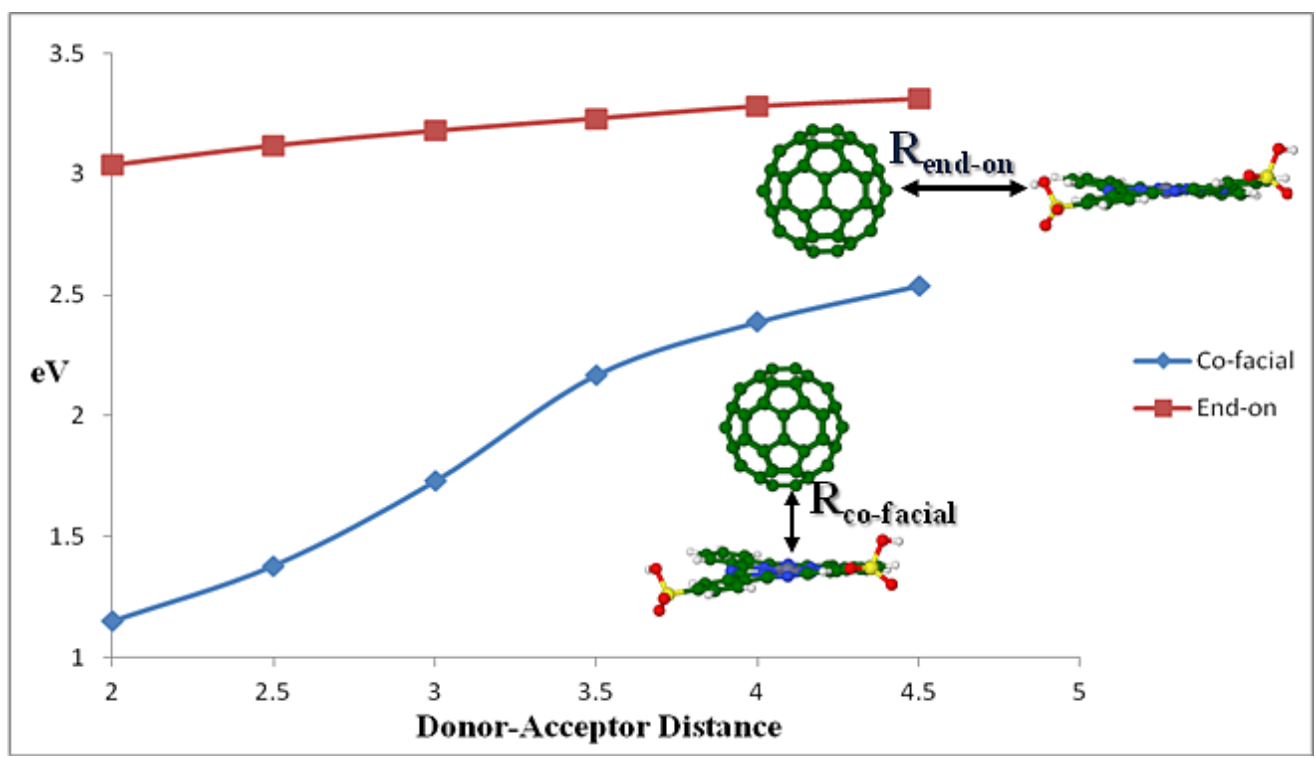

CONCLUSIONS